\documentclass[%
 reprint,
 amsmath,amssymb,
 aps,
]{revtex4-2}
\usepackage{graphicx}% Include figure files
\usepackage{dcolumn}% Align table columns on decimal point
\usepackage{bm}% bold math
\usepackage{tikz}
\usepackage{lipsum,adjustbox}
\usetikzlibrary{shapes.geometric, arrows,automata,positioning}

\tikzstyle{startstop} = [rectangle, rounded corners, minimum width=3cm, minimum height=1cm,text centered, text width=3cm, draw=black]

\tikzstyle{io} = [trapezium, trapezium left angle=70, trapezium right angle=110, minimum width=3cm, minimum height=1cm, text centered, draw=black, fill=blue!30]

\tikzstyle{process} = [rectangle, minimum width=3cm, minimum height=1cm, text centered, draw=black, fill=orange!30]
\tikzstyle{decision} = [diamond, minimum width=3cm, minimum height=1cm, text centered, draw=black]

\tikzstyle{arrow} = [thick,->,>=stealth]

\usepackage{circuitikz}

\newsavebox{\mycircuit}
\sbox{\mycircuit}{%
\begin{circuitikz}

    \node[inner sep=0pt] (FBM) at (-2,-0.5)
    {\includegraphics[width=1\textwidth]{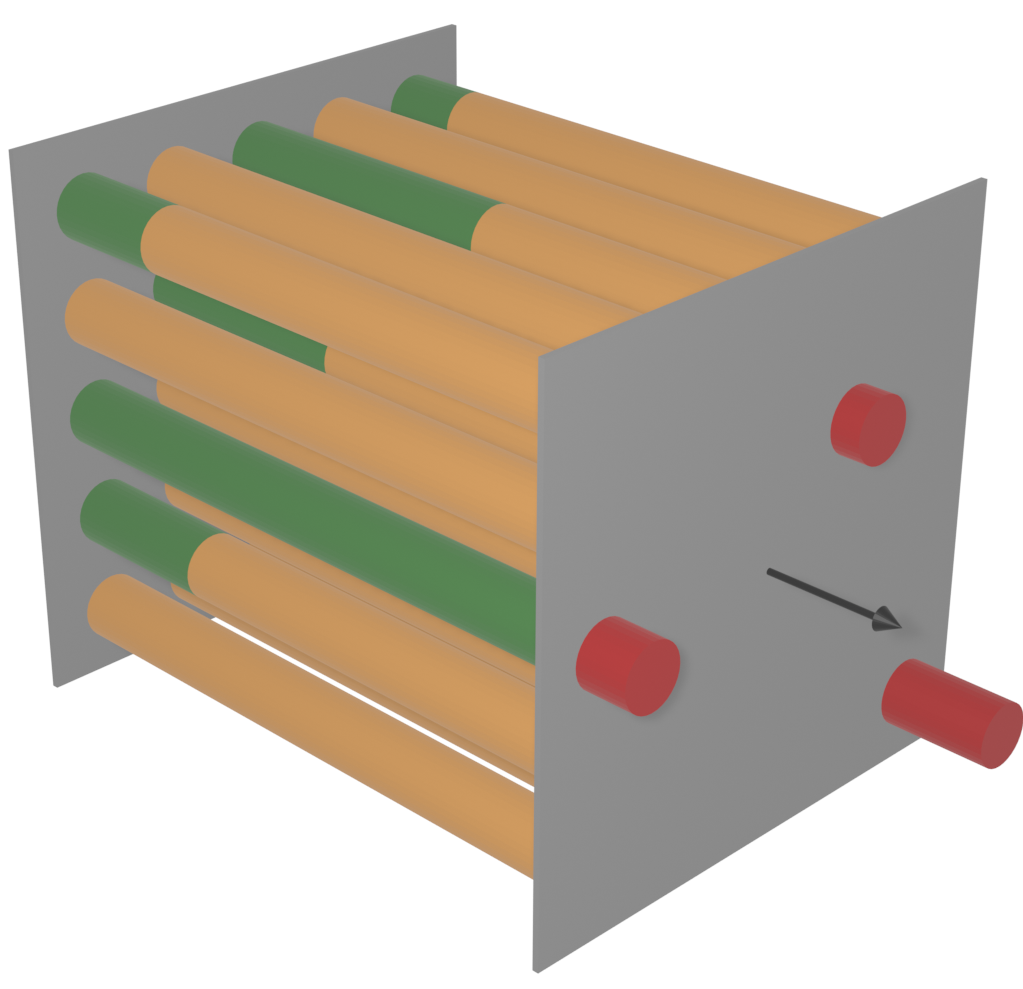}};
    \draw[ultra thick, ->] (-1,1.7) .. controls (0,4) and (1,4) .. (3,3);
    \draw[->] (3.5,2.9) to[short] node[anchor=north, yshift=0.5cm] {$\sigma_i$} (3.5,3);
    \draw[] (3.5,2.9) to[spring, l=$D(\varphi)$] (3.5,1);
    \draw[] (3.5,1) to[short] (2.5,1)
                              to[damper, l=$\tau$] (2.5,0)  
                              to[short] (3.5,0);
    \draw[] (3.5,1) to[short] (4.5,1)
                              to[spring, l=$J(\varphi)$] (4.5,0)
                              to[short] (3.5,0);

    \draw[] (3.5,0) to[thermistor,l=$\alpha$] (3.5,-2);
    \draw[thick] (3.5,-2) to[short] (3.9,-2);
    \draw[thick] (3.9,-2) to[short, l=$\varepsilon_i^S$] (3.9,-2.5);
    \draw[thick,-*] (3.9,-2.5) to[short] (3.6,-2.5);
    \draw[thick] (3.5,-2.2) to[short] (3.5,-2.7);
    \draw[->] (3.5,-2.5) to[short] node[anchor=south, yshift=-0.7cm] {$\sigma_i$} (3.5,-3.5);

    \draw[] (6,3) to[short] (6.4,3);
    \draw[] (6.2,3) to[short,l=$\varepsilon$] (6.2,-3.5);
    \draw[] (6,-3.5) to[short] (6.4,-3.5);
    \end{circuitikz}
    }

\begin{document}

\preprint{APS/123-QED}

\title{A Unifying Model for the Rheological Behavior of Hygro-responsive Materials}% Force line breaks with \\

\author{Júlio O. Amando de Barros}
 \email{jortiz@student.ethz.ch}%
\author{Falk K. Wittel}
\email{fwittel@ethz.ch}
\affiliation{%
 ETH Zurich, Institute for Building Materials}%

\date{\today}

\begin{abstract}
Hygro-responsive materials exhibit a complex structure-to-property relationship. The interactions of water within these materials under varying hygric and mechanical loads play a crucial role in their macroscopic deformation and final application. While multiple models are available in literature, many lack a comprehensive physical understanding of these phenomena. In this study, we introduce a novel Stick-Slip Fiber Bundle Model that captures the fundamental behaviors of hygro-responsive materials. We incorporate moisture-dependent elements and rules governing the initiation and relaxation of slip strains as well as failure to the statistical approach offered by Fiber Bundle Models. The additional features are based on well-founded interpretations of the structure-to-property relationship in cellulosic materials. Slip strains are triggered by changes in load and moisture, as well as by creep deformations. When subjected to moisture cycles, the model accumulates slip strains, resulting in mechanosorptive behavior. When the load is removed, slip strains are partially relaxed, and subsequent moisture cycles trigger further relaxation, as expected from observations with mechanosorptive material. Importantly, these slip strains are not considered plastic strains; instead, they are unified, non-linear frozen strains, activated by various stimuli. Failure of fibers is defined by a critical number of slip events allowing for an integrated simulation from intact, via damaged, failed states. We investigate the transition between these regimes upon changes in the hygric and mechanical loading history for relevant parameter ranges. Our enhanced Stick-Slip Fiber Bundle Model increases the understanding of the intricate behavior of hygro-responsive materials and contributes to a more robust framework for analyzing and interpreting their properties.
\end{abstract}

\maketitle

\section{\label{sec:Intro}Introduction}
Hygro-responsive materials behave in a non-trivial rheological way when mechanical load and moisture changes act simultaneously.
Current models decompose the behavior into distinct phenomena: elasticity, viscoelasticity, plasticity, hygro-expansion, and mechanosorption~\cite{ranta1975viscoelasticity,fortino20093d,hassani2015rheological}. Each is considered by an independent rheological element, whose response is superimposed to predict the total strain. Such an approach relies on fitting models to data to determine constitutive parameters for predicting the material response to various stimuli, but no physical insight is gained this way. When thinking about the microscale of materials with a disorder, it becomes evident that the rheological behavior emerges from the dynamic moisture-dependent interaction of constituent components. Interactions can be determined by hydrogen bonds where water molecules can link to, rendering the interaction of constituents, and thus resulting in moisture-dependent behavior. An example of a natural material is wood, where cellulose microfibrils interact via a hemicellulose and lignin-rich matrix~\cite{boyd1982anatomical,dong2010fibre,engelund2013critical,stevanic2020molecular}. The emerging mechano-sorptive, viso-elastic, or plastic strains, contributing to the overall macroscopic strain, can be associated with inter-fibrillar displacements \cite{alfthan2004micro}. This leads to a unifying perspective, where one type of micromechanism manifests in different strain contributions. For wood fibers, the identified mechanism is stick-slip movement within microfibril aggregates (MFA)~\cite{ZHANG2021117682} or between those, resulting in permanent or temporary strains that we will call slip strains in the following.

%Why should we care?
The technological importance of hygro-responsive behavior in natural fibrous materials such as wood, flax, linen, wool, synthetic fibers, or even concrete is evident ~\cite{pickett1942effect,mackay1959effect,armstrong1961influence,byrd1972effect,wang1990transient,wang1993effects,olsson2007mechano}. Mechanosorptive creep, emerging from moisture changes under mechanical load, is among the least understood observations \cite{engelund2013critical}. Early works already speculated on the role of the mobility of hydrogen bonds within wood fibrils during moistening and drying events as the source of mechanosorption~\cite{armstrong1960effect,nordon1962some,gibson1965creep}. This was complemented by explanations regarding the emergence of transient stresses due to water dispersion during absorption by \citet{mackay1959effect,pickett1942effect}. Recently, infrared spectroscopy of wood mechanosorption proved the mobility of hydrogen bonds resulting in slipping processes~\cite{stevanic2020molecular}. This way, moisture-rate-dependent creep strains and frozen strains can be qualitatively explained. Unfortunately, related models are rarely comprehensive, as pointed out by~\citet{alfthan2004micro}. The main criticism addresses the lack of generality beyond specific setups or materials~\cite{hoffmeyer1989mechano,engelund2011modelling}, as well as the utilization of mechanisms without consistent physical background~\cite{van1989theoretical,padanyi1993physical,gril2015modelling}. Fiber bundle models (FBM) are a class of models from statistical mechanics, that greatly contributed to a comprehensive understanding of complex structure-property relations. Dating back to the 1920s by~\citet{peirce1926tensile}, these models were extensively adopted to study the damage evolution, fracture, or size effects in materials with disorder since then~\cite{phoenix_statistical_2000,hansen2015fiber}. Recently, a fiber-bundle-based modeling approach was proposed  by~\citet{halasz2009fiber}, where stick-slip behavior was included on fiber-bundles (SS-FBM). The model proved to capture the concept of conformational changes triggered by external loads.

%3rd: What do we do about the problem?
In this work, we add the basic effects of the moisture dependence to the stick-slip FBM, by introducing moisture-dependent behavior for constituting elements and their interactions. The identified rules are inspired by experimental and numerical findings from wood cell wall mechanics \cite{stevanic2020molecular,zhang2021hydrogen}. In detail, the main modifications introduced are moisture-dependent constitutive parameters such as compliance and slip thresholds, the possibility for reverse slip and recovery, as well as the definition of a resulting internal history variable. To facilitate explanations and discussions, we utilize the association with microfibrils and hydrogen bonds, present in wooden fibers, even though our work goes beyond this mindset. The manuscript is organized as follows: In the materials and methods section, we first review the main phenomenological observations of mechanosorptive creep and the consequences of the model assumptions. Then, we give a complete description of the SS-FBM and its extensions accompanied by a description of important implementation details. In the results section, we first study the generic model behavior and the sensitivity on parameters. We explore the parameter space by simulating reasonable moisture and mechanical loading situations. Furthermore, we study the consequences of the evolution of the fine structure of creep and damage inside the bundle on the macroscopic observations. 
Additionally, the effects of model parameters on the system's spontaneous and creep failure are investigated. Finally, we conclude that the macroscopic mechanosorptive, visco-elastic, elastic, and plastic behavior emerge from the same underlying mechanism and can be understood and modeled in a unified way.

%M&M%%%%M&M%%%%M&M%%%%M&M%%%%M&M%%%%M&M%%%%M&M%%%%M&M%%%%M&M%%%%M&M%%%%M&M%%%%M&M%%%
\section{Materials and Methods}\label{sec:mandm}
To prepare the ground for the interpretation of the hygro-responsive model, we first discuss the main observed phenomena in the framework of cell wall mechanics (see Sec.~\ref{Sec:experimental}). We then address the required extensions to the stick-slip Fiber Bundle Model (FBM) (see Sec.~\ref{Sec:SS-FBM}), before we give the mathematical formulation as well as some implementation details and simulation parameters (see Secs.~\ref{Sec:FBM-math},\ref{Sec:FBM-code}).
%%%%%%%
\subsection{Experimental observations and interpretation}\label{Sec:experimental}
\begin{figure}
\begin{center}
\resizebox{\columnwidth}{0.5\columnwidth}{
\begin{tikzpicture}[]
\large
\draw[->] (0,-7) to[short] node[anchor=south, yshift=-0.7cm,xshift=-85] {$Time$} (11,-7);
    \draw[->] (0,-7) to[short] node[anchor=south, yshift=-1.6cm, xshift=-7] {$\varepsilon$} (0,-1.7);
    \draw[->, blue] (11,-7) to[short] node[rotate=-90, xshift=1.5cm, yshift= 0.5cm] {$RH(\%)$} (11,-1.7);
    \draw[->, red] (12.3,-7) to[short] node[rotate=-90, xshift=1.5cm, yshift= 0.5cm] {$\sigma$} (12.3,-1.7);

    %MOISTURE 
    
    \draw[dashed,blue] (0,-6) to[short] (0.5,-6);
    \draw[dashed,blue] (0.5,-6) to[short] (0.5,-5);
    \draw[dashed,blue] (0.5,-5) to[short] (1.25,-5);
    \draw[dashed,blue] (1.25,-5) to[short] (1.25,-6);
    
    \draw[dashed,blue] (1.25,-6) to[short] (3,-6);
    \draw[dashed,blue] (3,-6) to[short] (3,-5);
    \draw[dashed,blue] (3,-5) to[short] (3.75,-5);
    \draw[dashed,blue] (3.75,-5) to[short] (3.75,-6);
    
    \draw[dashed,blue] (3.75,-6) to[short] (4.5,-6);
    \draw[dashed,blue] (4.5,-6) to[short] (4.5,-5);
    \draw[dashed,blue] (4.5,-5) to[short] (5.25,-5);
    \draw[dashed,blue] (5.25,-5) to[short] (5.25,-6);
    
    \draw[dashed,blue] (5.25,-6) to[short] (7,-6);
    \draw[dashed,blue] (7,-6) to[short] (7,-5);
    \draw[dashed,blue] (7,-5) to[short] (7.75,-5);
    \draw[dashed,blue] (7.75,-5) to[short] (7.75,-6);
    
    \draw[dashed,blue] (7.75,-6) to[short] (8.5,-6);
    \draw[dashed,blue] (8.5,-6) to[short] (8.5,-5);
    \draw[dashed,blue] (8.5,-5) to[short] (9.25,-5);
    \draw[dashed,blue] (9.25,-5) to[short] (9.25,-6);
    
    \draw[dashed,blue] (9.25,-6) to[short] (11,-6);

    % Load
    
    \draw[dashed,red] (2,-7) to[short] (2,-5.2);
    \draw[dashed,red] (2,-5.2) to[short] (6,-5.2);
    \draw[dashed,red] (6,-5.2) to[short] (6,-7);

    % Strain
    
    \node[text width=0.1cm] at (0.8,-6.3) {I};

    \draw[thick] (0.5,-7) to[short] (0.5,-6.7);
    \draw[thick] (0.5,-6.7) to[short] (1.25,-6.7);
    \draw[thick] (1.25,-6.7) to[short] (1.25,-7);

    \node[text width=0.1cm] at (1.7,-4.6) {II};

    \draw[thick] (2,-7) to[short] (2,-4.5);

    % \node[text width=0.1cm] at (2,-4) {VE};
    
    \draw[thick] (2,-4.5) .. controls (2.5,-4.1) and (2.25,-4.2)  .. (3,-4);

    % \draw[thin,dotted] (2.75,-2.5) rectangle ++(3.4,-1.7);
    \node[text width=0.1cm] at (4.3,-2.3) {III};

    \draw[thick] (3,-4) to[short] (3,-3.5);
    
    \draw[thick] (3,-3.5) .. controls (3.25,-2.95) and (3.5,-2.85)  .. (3.75,-2.9);

    \draw[thick] (3.75,-2.9) to[short] (3.75,-3.4);        

    \draw[thick] (3.75,-3.4) .. controls (4,-3.65) and (4.25,-3.75)  .. (4.5,-3.7);

    \draw[thick] (4.5,-3.7) to[short] (4.5,-3.2);
    
    \draw[thick] (4.5,-3.2) .. controls (4.75,-2.8) and (5,-2.7)  .. (5.25,-2.7);

    \draw[thick] (5.25,-2.7) to[short] (5.25,-3.2);

    \draw[thick] (5.25,-3.2) .. controls (5.5,-3.35) and (5.75,-3.39)  .. (6,-3.4);

    \node[text width=0.1cm] at (6.1,-4.6) {IV};

    \draw[thick] (6,-3.4) to[short] (6,-6.2);

    % \node[text width=0.1cm] at (6.1,-6.6) {VE};
    
    \draw[thick] (6,-6.2) .. controls (6.25,-6.4) and (6.5,-6.45)  .. (7,-6.5);

    % \node[text width=0.1cm] at (7.3,-6) {H};

    \draw[thick] (7,-6.5) to[short] (7,-6.2);
    \draw[thick] (7,-6.2) to[short] (7.75,-6.2);
    \draw[thick] (7.75,-6.2) to[short] (7.75,-6.7);

    \draw[] (7.75,-6.7) to[short] (8.5,-6.7);

    \node[text width=0.1cm] at (8,-6.3) {V};
    
    \draw[thick] (8.5,-6.7) to[short] (8.5,-6.5);
    \draw[thick] (8.5,-6.5) to[short] (9.25,-6.5);
    \draw[thick] (9.25,-6.5) to[short] (9.25,-6.9);
    
    \draw[thick] (9.25,-6.9) to[short] (10.5,-6.9);

    % \draw[thin,dotted] (8.2,-7) rectangle ++(9,-6.3);

\end{tikzpicture}
}
\caption{\label{fig:Strain} Main features observed in hygro-responsive materials. The strain ($\varepsilon$) is presented in black, while the load ($\sigma$) and relative humidity (RH) are represented by red, resp. blue dashed lines. Letters indicate different features like H for hygroexpansion, E for elasticity, VE for viscoelasticity, and MS for mechanosorption.}
\end{center}
\end{figure}
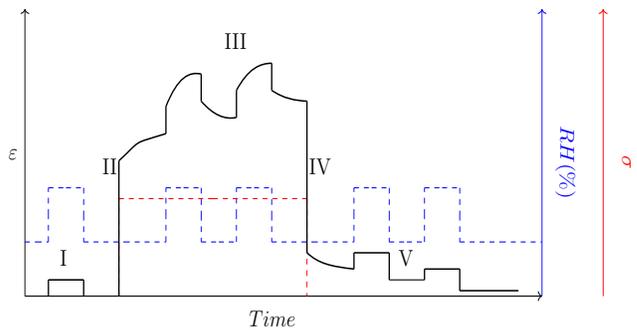

Combined mechanical and hygric stimuli both manifest in a rheological behavior composed of scleronomous deformation mechanisms such as elastic, hygro-expansive, plastic, but also mechano-sorptive strains, and rheonomous ones like the visco-elastic response. This combination is exemplified in  Fig.~\ref{fig:Strain}, where alternating dry and moist states are superimposed with a mechanical load-interval: Without external load, moisture variations result in free swelling or shrinkage hygro-expansive strain~\cite{skaar1988hygroexpansion} (see Fig.~\ref{fig:Strain}-I). When ramping up the load, instantaneous elastic deformation occurs, and depending on the loading degree, significant plastic strains are observed as well~\cite{green1999mechanical}(see Fig.~\ref{fig:Strain}-II). At the same time, visco-elastic creep strains start to evolve~\cite{gibson1965creep}. If we now change the moisture again, the elastic and visco-elastic strains would adapt through moisture-dependent elastic and creep material compliances. However, one observes an additional strain component that is triggered by moisture changes under load, namely the mechano-sorptive strain (see Fig.~\ref{fig:Strain}-III), interestingly under moistening and demoistening~\cite{armstrong1960effect}. Elastic and viscoelastic strains relax after unloading, but a remaining strain is observed (see Fig.~\ref{fig:Strain}-IV). The remaining frozen strain can relax when moisture alternates in the unloaded system (see Fig.~\ref{fig:Strain}-V). Microstructural features of the cell wall determine this complicated hygro-mechanical behavior. Therefore, we briefly review the dominating structure-property relations. 
 
\begin{figure*}
\centering
\includegraphics[width=\textwidth]{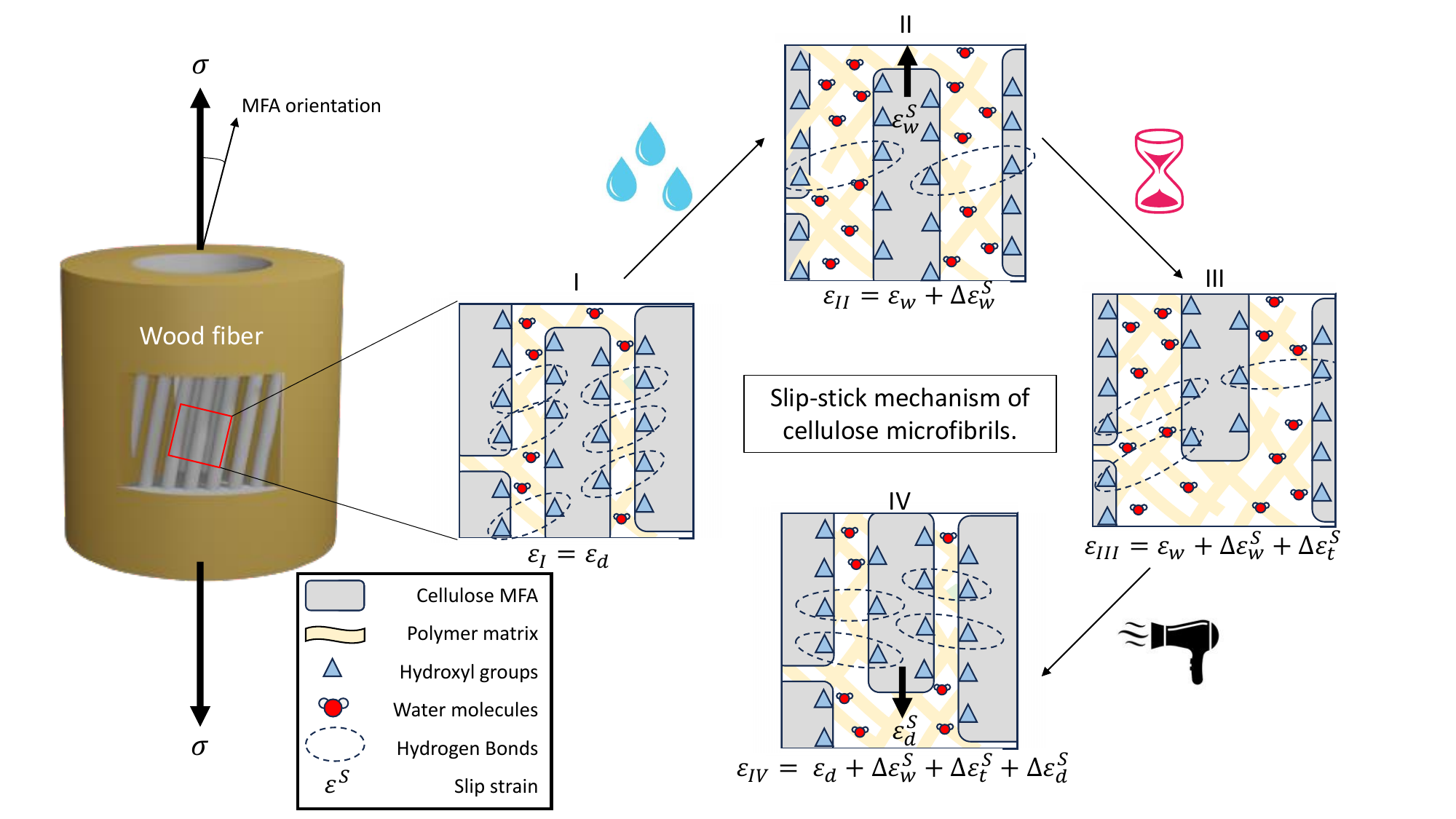}
\caption{Micromechanical interpretation of stick-slip dynamics on wood fibers. We present a wood fiber on the left with its microfibril aggregates exposed. Following to the right, the aggregates' interactions and their behavior under mechanical and moisture stimuli are presented. $\varepsilon_d$ and $\varepsilon_w$ are the non-slip strain at the dry state and wet state, respectively, while $\varepsilon^S_d$, $\varepsilon^S_w$ and $\varepsilon^S_t$ are the slip strains respective to drying, wetting, and time evolution.}
\label{fig:mechanism_representation}
\end{figure*}

The cell wall is at the core of all accepted explanations for the hygro-mechanical behavior, and in particular its chemical and ultra-structural composition~\cite{salmen2009cell}. In a nutshell, cellulose microfibril aggregates are embedded in a polymer matrix of hemicelluloses and lignin (Fig.~\ref{fig:mechanism_representation}-I)~\cite{brandstrom2001micro}. Microfibrils are about 15 to 20 times stiffer than the embedding matrix but only in the fibril direction at an angle to the longitudinal fiber orientation, called the microfibril angle, which is not relevant for this work. The important macro bio-polymer molecules and microfibril aggregates interact strongly through hydrogen bonds. By increasing the moisture content, water molecules will compete for the superficial hydrogen bonds of MFAs (Fig.~\ref{fig:mechanism_representation}-II)~\cite{engelund2013critical,salmen2018effect}. This will reduce the bonding density, result in swelling in the transverse direction, changes in overall compliance, and can result in relative displacement of neighboring aggregates, called slips in the following~\cite{stevanic2020molecular}. After a slip, a microfibril aggregate will form new hydrogen bonds with new neighbors (Fig.~\ref{fig:mechanism_representation}-III). Slips can trigger slip avalanches due to load redistribution, resulting in complex slip dynamics. 

In this manuscript, we demonstrate that moisture and history-dependent stick-slip rules, applied to uni-axial fiber bundle models (FBM), have the potential to reproduce and explain the entire range of observed phenomena in hygro-mechanical loading in a unified way.
%%%%%%%%%%%%%%%%%%%%%%%%%
\subsection{Extensions to the Fiber Bundle Model (FBM) with stick-slip}\label{Sec:SS-FBM}
A FBM is composed of a grillage of rheological elements with disorder, that interact mechanically when loaded. In its simplest form, elements would be Hooke's bodies, with their load being redistributed to all other intact elements of the bundle (democratic load sharing) upon reaching a failure threshold~\cite{daniels1945statistical}. For materials failure, identical constitutive parameters but Weibull-distributed~\cite{weibull1951statistical} breaking thresholds proved suitable~\cite{kun2000damage}. A wide variety of constitutive behaviors can be realized by altering the load redistribution strategy from democratic or global to local, meaning the nearest intact neighbors will experience more shared load than distant ones~\cite{zhang1996failure}. Time-dependent behavior can be realized by introducing different rheological models for the elements, like Kelvin-Voigt bodies~\cite{hidalgo2002creep}. Alternatively, a visco-plastic behavior of the system can be realized by allowing fibers to debond when reaching a threshold and reattach within the bundle, called stick-slip fiber-bundle model (SS-FBM) as proposed by ~\citet{halasz2009fiber}. The last approach suits materials such as cellulose materials, where conformational changes occur under mechanical and hygric stimuli. However, the SS-FBM does not have features of hygro-responsive materials. 

% Each feature added to the proposed SS-FBM takes different roles in the mechanosorptive dynamics.
To include the conceptual understanding for simulating the observed behavior of mechanosorption, three major adaptations to the SS-FBM are proposed:
\begin{enumerate}
    \item Moisture-dependent rheological elements;\label{MD} Elastic and viscoelastic compliances increase with moisture~\cite{skaar1988hygroexpansion} due to the previously mentioned decrease in hydrogen bond density~\cite{peirce192916}. 
    \item Multi-directional slip, extending the capabilities of SS-FBM to unloading and reverse loading, motivated by the identical shear-lag mechanism under tension and compression inside the bundle~\cite{dwaikat2013predicting}.\label{SB} 
    \item Moisture and history-dependent slip threshold values to consider the effects of moisture-induced reduction of hydrogen bonds and increasing mismatch in bonding-pair sites with the relative displacement of MFAs due to slips. Additionally, the initial positions are considered more stable configurations, and it will be less costly to slip back. \label{TH}
\end{enumerate}

In the following, we give the mathematical formulation of the FBMs and describe our implementation.
%%%%%%%%%%%%%%%%%%%%%%%%%%%%%%%%%%%%%%%%%%%%%%%%%%%%%%%%%%%%
\subsection{Mathematical Formulation}\label{Sec:FBM-math}
FBMs are an assembly of $N$ parallel fibers with identical constitutive behavior except for the failure threshold. The bundle reacts with a strain $\varepsilon$ as a response to an external load stimulus $\sigma$. All $i=1,2,3,..., N$ fibers share the external load equally up to the failure of the first fiber $i$. Once the failure threshold of fiber $i$ is reached, it fails and its load $\sigma_i$ is redistributed to all intact fibers. In an SS-FBM, stick-slip motion is added to the model. At failure or slip of fiber $i$, when reaching the slip threshold strain $\varepsilon_i^{th}$, its load is set to zero and $\varepsilon_i^{th}$ is added to its permanent strain $\varepsilon_i^S$. Next, the fiber is reattached to the bundle and can be further loaded when its previous load gets redistributed to all intact fibers. A fiber can slip $k^f$ times, with the same or an altered slip threshold $\varepsilon_i^{th}(k)$, where $k=0,1,2,..., k^f$. After slipping $k^f$ times, the fiber can hold no further load and is considered as broken or mechanically detached.

\begin{figure}
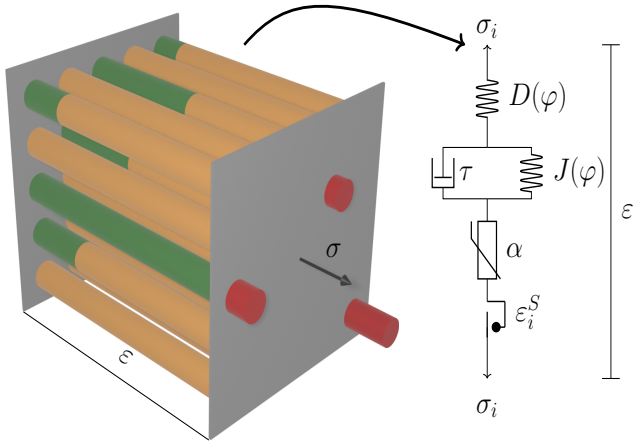

\resizebox{\columnwidth}{0.7\columnwidth}{
\include{Images/FBM_pic}
}
\caption{Sketch of the FBM with green representing slip, yellow tensile, and red compressive strains. On the right is the rheological representation of a single fiber model with elastic, viscoelastic, hygro-expansive, and slip element in series. The total strain $\varepsilon$ is present in both representations.}
\label{fig:rheologic}
\end{figure}

Our fibers are composed of a moisture-dependent linear elastic element with compliance $D(\varphi)$, where $\varphi$ is the moisture content, and elastic strain $\varepsilon^E_i$, a moisture-dependent linear viscoelastic element with compliance $J(\varphi)$, characteristic time $\tau(\varphi)$, and the viscoelastic strain $\varepsilon^{VE}_i$. Hygro-expansive strain $\varepsilon^H_i$ is calculated proportional to $\varphi$ by an expansion factor $\alpha$. The slip element is defined to represent the fiber's cumulative slip strain $\varepsilon_i^S=\sum_{j=1}^k \varepsilon_{i}^{th}(k,\varphi,\beta)$, already including the additional slip threshold dependence on $\varphi$ and on the slip direction $\beta$. The rheologic model is sketched in Fig. \ref{fig:rheologic}.

In a fiber $i$, the force $F_i$ is identical for all rheological elements in series, such as the Hookean body and the Kelvin body, namely
\begin{equation} 
F_i=A_iD(\varphi)^{-1}\varepsilon_i^{E}=A_iJ(\varphi)^{-1}\big(\varepsilon_i^{VE}-\tau \dot\varepsilon_i^{VE}),\label{1_fiber_ve}
\end{equation}
with the unit area $A_i=1$. The macroscopic stress $\sigma$ is given by the sum of all forces divided by the total area, namely $\sigma= \sum^N F_i/A_i =  \sum^N F_i/N$. The macroscopic strain is identical for all $N$ fibers with $\varepsilon=\varepsilon^E_i+\varepsilon^{VE}_i+\varepsilon^{H}_i+\varepsilon^S_i$, but of course, individual strain contributions may vary for each fiber.

Analytical solutions could be found for some special cases in SS-FMBs \cite{halasz2009fiber,halasz2021stick}, and viscoelastic FBMs \cite{hidalgo2002creep}. In the framework of our work, however, material nonlinearity, local interaction rules, and alterations of the threshold distribution require a numeric, incremental solution. We calculate the systems response for given load and moisture profiles, $\sigma(t),\varphi(t)$. The different contributions to the total strain of a fiber that equals the one of the bundle $\varepsilon_i=\varepsilon$ are given in incremental form by
    \begin{align}
        \Delta\varepsilon^E_i (\varphi)&=D(\varphi)\Delta\sigma_i,\label{elastic}\\
        \Delta\varepsilon^{VE}_i(\varphi,t)&=J(\varphi)\sigma_i\big(1-e^{-(t\Delta t/\tau}),\label{viscoelastic}\\
        \Delta\varepsilon^{H}_i(\varphi)&=\alpha\Delta\varphi,\label{hygro}\\
        \varepsilon^S_i(k)&=\sum_{j=1}^k \varepsilon_i^{th}(j),\label{slip}
    \end{align}
and the macroscopic strain by
    \begin{equation}
    \varepsilon=\frac{1}{N}\big(D(\varphi)\sigma +\varepsilon^{VE} +\varepsilon^{H}+\varepsilon^S \big)
    \label{epsilonmacro}
    \end{equation}
 $\varepsilon^{VE}$, $\varepsilon^{H}$, and $\varepsilon^{S}$ are the sum of viscoelastic, hygroexpansive, and slip strains, respectively, over all fibers. The sum of elastic strains, $\varepsilon^{E}$, is substituted by $D(\varphi)\sigma$ since all fibers have the same compliance and $\sigma$ is the sum of all forces. 

The slip condition of each fiber is obtained by 
\begin{equation} \label{slip_cond}
\varepsilon-\varepsilon_i^S\geq\varepsilon_{i}^{th}(k,\varphi,\beta).    
\end{equation}
with
\begin{equation} \label{slipthesredcond}
\varepsilon_{i}^{th}(k,\varphi,\beta)=f(k)\Gamma_{\varphi}\Gamma_{\beta}\varepsilon_{i}^{th,0}.
\end{equation}
$f(k)$ is a discrete decreasing function of $k$, $\Gamma_{\varphi}$ and $\Gamma_{\beta}$ are influence factors for the moisture and slip direction respectively, and $\varepsilon_{i}^{th,0}$ denotes the initial slip threshold sampled from a Weibull distribution (values see Tab.~\ref{tab:tab_par}).
%%%%%%%%%%%%%%%%%%%%
\subsection{Model Implementation and Parameter Space}\label{Sec:FBM-code}
\begin{figure}
        \centering
        \includegraphics[width=1\columnwidth]{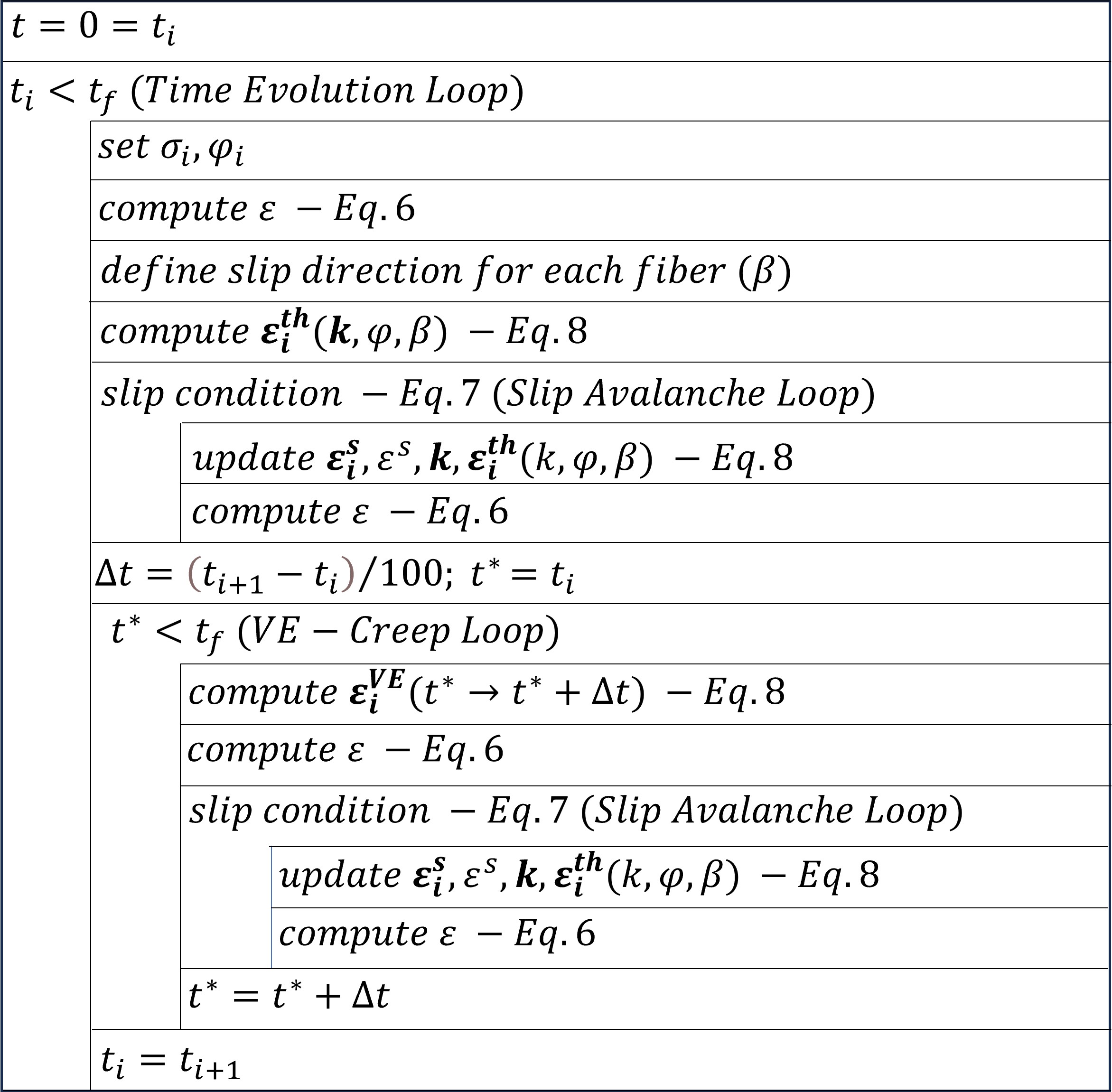}
        \caption{Flowchart of implemented code.}
        \label{fig:chart}
    \end{figure}

To allow for the wide exploration of the parameter space, we make a flexible implementation of the SS-FBM, schematized in Fig.~\ref{fig:chart}. We follow a given time evolution of stress $\sigma(t)$ and moisture $\varphi(t)$ from $t_0$ to $t_{final}$ (see Fig.\ref{fig:Strain}) in a step-wise fashion. Hence, we assume constant boundary conditions for the duration of one step.

First, we solve the scleronomic response of the system upon an external load or moisture increment, due to all triggered slip events (see slip avalanche loop in Fig.~\ref{fig:chart}). To ensure that our solution stays on the non-linear solution path, we continuously identify the most critical fiber and slip it, followed by an update until no more slip events occur. 

The rheonomous response within a time step is integrated for each fiber in 100 integration steps, followed by the slip avalanche loop after each integration step (see VE-creep loop in Fig.~\ref{fig:chart}). During the simulation, we record the evolution of all relevant state variables for further analysis.
    
The FMBs in this study are formed by a square lattice with $200\times 200$ fibers. For simplicity, we limit ourselves to two extreme relative humidity values called $RH_d$ and $RH_w$ for the dry and wet state. In the different states, the elastic compliance of the wet state $D_w$ is larger than the dry one $D_d$, and for the viscoelastic compliances $J_d<J_w$ holds. Values are summarized in Tab.~\ref{tab:tab_par}. Note that $D_d$ is taken as unity, while all other compliances relate to it. The dry FBM without slip ($k^f=1$), has a strength of $\sigma_c$ and a failure strain $\varepsilon_c$, and is taken to normalize the applied load $\sigma$ or macrostrain $\varepsilon$. To relate simulation time $t$ to the system behavior, it is normalized by the characteristic time of the Kelvin-Voigt element $\tau$. A typical wet time interval is chosen to be $\Delta t_w=10\tau$.

The slip thresholds $\varepsilon_i^{th}$ are sampled from a Weibull distribution with a constant shape parameter $m$, but moisture and history-dependent scale parameter $\lambda(\varphi,k)=\Gamma_\varphi \Gamma_\beta e^{-k/d}$. $d$ is a chosen decay parameter, $\Gamma_\varphi$ a factor for the moisture influence while reverse loading further reduces $\lambda$ by the rescaling factor $\Gamma_\beta$ (see Tab.~\ref{tab:tab_par}). Note that fibers are only allowed to slip $k^f$ times, which corresponds to $e^{-k^f/d}=0.3=\varkappa^f$. In this way, the failure criterion has the same physical meaning (weakening limit) for any value of $d$.

\begin{table}[htb]
\caption{\label{tab:tab_par}%
Set of simulation parameters and variation ranges.}
\begin{tabular}{ccl}
Parameter & Normal value & Reference \\
\hline \hline
\multicolumn{3}{l}{Bundle and simulation properties:} \tabularnewline
N & 40000 & number of fibers \\
$\sigma_c$ & 2.27 & FBM dry strength\\
$\sigma/\sigma_c$ & 0.5 & normalized applied load\\
$\Delta t_w$ & $10\tau$ & time interval for wet states\\
\multicolumn{3}{l}{Fiber mechanical properties:} \tabularnewline
$D_d$ & 1 & dry state elastic compliance\\
$D_w$ & 1.3 & wet state elastic compliance\\
$J_d$ & 0.2 & dry state creep compliance\\
$J_w$ & 0.4 & wet state creep compliance\\
$\tau$ &  1  & creep characteristic time\\
$\alpha$ & 0.5 & swelling coefficient\\
\multicolumn{3}{l}{Threshold properties:} \tabularnewline
$m$ & 2 & Weibull shape parameter \\
$\lambda$ & 7 & Weibull scale parameter \\
$f(k)$ & $e^{-k/d}$ & weakening function \\
$d$ & 11 & decay parameter of $f(k)$ \\
$\varkappa^f$ & 0.3 & failure limit\\
$k^f$ & 13 & calculated slip number \\
$\Gamma_m$ & 0.7 & wet state rescaling factor \\
$\Gamma_\beta$ & 0.7 & reverse slip rescaling factor \\
\multicolumn{3}{l}{Ranges used for simulations:} \tabularnewline
$D_w/D_d$ & $1.05-1.3$ & Sec.~\ref{Sec: Res: Single}\\
$D_w/D_d$ & $1.05-2.0$ & Sec.~\ref{Sec: Res: fail}\\
$d$  & $5-55$ & Sec-~\ref{Sec: Res: fail}\\
$\varkappa^f$  & $0.005-0.30$ & Sec-~\ref{Sec: Res: fail}\\
\end{tabular}
\end{table}

%%%%%%%%%%%%%%%
\section{Results and Discussion}\label{Sec. Results}
The system behavior is discussed and analyzed in three steps: First, we investigate the behavior and development of slip strains for a single moisture cycle (MC) in mechanical loading and unloading. Then, we study the strain evolution with repetitive MCs, showing the emergence of mechanosorptive strains. Finally, we analyze the macroscopic failure.
%%%%%%%%%%%%%%%%
\subsection{Analysis of SS-FBM behavior for a single moisture-cycle}\label{Sec: Res: Single}
\begin{figure}
    \centering
    \includegraphics[width=\columnwidth]{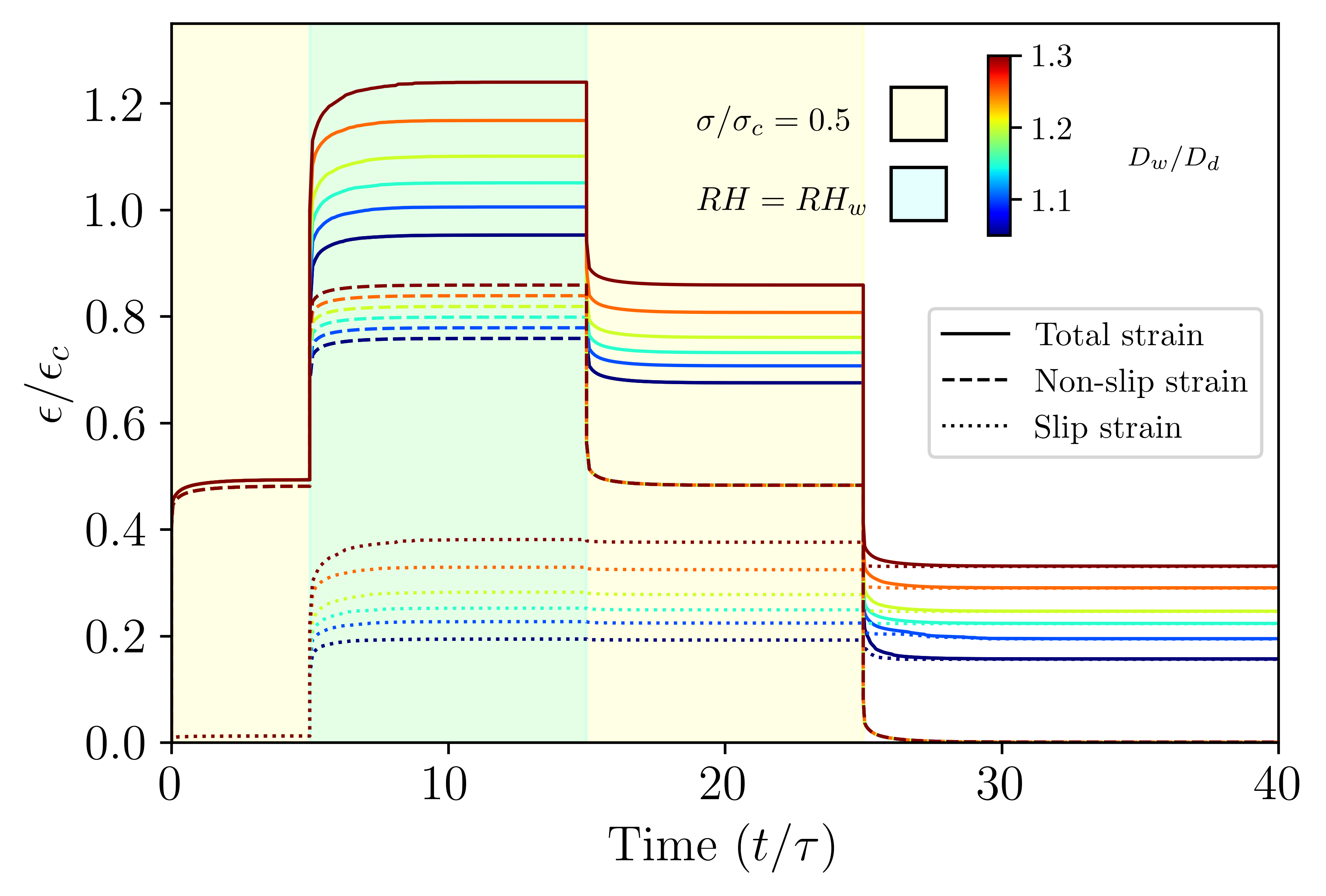}
    \caption{SS-FBM normalized strain over the normalized time. Solid lines represent the normalized strain $\varepsilon/\varepsilon_c$, while dashed lines represent the normalized strain without slip $(\varepsilon-\langle\varepsilon^S\rangle )/\varepsilon_c$. Line colors resemble softening ratios $D_w/D_d$.}
    \label{fig:1cycle}
\end{figure}

To gain insight into the role of the slip dynamics on the bundle behavior, we first analyze a single moisture cycle within a load interval. While the mechanical load $\sigma/\sigma_c$ acts in the interval $t=[0,25]\tau$, the moisture is elevated to the wet state $RH_w$ during the interval $t=[5,15]\tau$, while the rest of the time it remains at $RH_d$ (see Fig.~\ref{fig:1cycle}). We monitor the evolution of the normalized macrostrain $\varepsilon/\varepsilon_c$ (solid lines), and decompose it into the normalized non-slip strain $(\varepsilon-\langle\varepsilon^S\rangle )/\varepsilon_c$ (dashed lines) and the normalized slip strain $\langle\varepsilon^S\rangle /\varepsilon_c$. To study the effect of the moisture softening on the slip dynamics, we calculate the evolution for various softening ratios $D_w/D_d$ from $1.05$ to $1.3$.

At $t=0$, the load $\sigma/\sigma_c$ is applied to the bundle in the dry state $RH_d$, which results in elastic strains, as well as the triggering of first slip events. During the interval $t=[0,5]\tau$, visco-elastic creep strain develops, and increased slip strains are observed as more fibers reach the threshold strain. Since the compliances $D_d,J_d$ are identical for all systems, the observed deformations are consequently identical. At $t/\tau=5$, we increase the relative humidity to the wet state, $RH_w$, which uniformly expands the bundle due to hygric swelling, as well as due to the compliance reduction to $D_w$, caused by the moisture softening. The increased strain and the reduced slip thresholds of the wet state now promote further fiber slipping. The equally spaced dashed lines show the chosen regular spacing between values of $D_w/D_d$, while the dotted lines show the slip strain acquired by each system. In general, the higher the increase of non-slip strains due to moisture increase, the higher the triggered slip strains. The increased creep compliance $J_w$ at wet state also clearly manifests in the strain evolution with higher creep strains. 

At $t=15\tau$, the system returns to its dry state, $RH_d$, and the entire bundle shrinks due to hygric shrinking as well as moisture hardening, and the dashed lines for the behavior without slip collapse again. At the fiber level, the drying event results in the compression of the fibers with high accumulated slip strain in the vicinity of fibers with lower slip number (as represented in red in Fig.~\ref{fig:rheologic}). Force balance requires that fibers under tension are additionally elongated to cope with the compressive stress, resulting in residual stress. This, in combination with the return to the higher slip thresholds for the dry state but compensated by the threshold reduction for a reverse slip by factor $\Gamma_\beta$, can trigger a reverse slip. Additionally, increased slip numbers $k_i$ further simplify slips due to the weakening function. As a consequence, residual stresses inside the bundle are relaxed only due to reverse slip events. This reduces the residual tensile load on the fibers under tension, decreasing the macroscopic strain $\varepsilon$. Reduced $\varepsilon$ again adds compressive stress on the fibers in compression, which then relaxes by further slips when their strains are below the slip thresholds, which can further decrease $\varepsilon$, and so on. This is superimposed by the strain recovery of the visco-elastic Kelvin-Voigt element, whose equilibrium strain is reduced when decreasing compliance $J_d$. At $t=25\tau$, the load $\sigma/\sigma_c$ is removed, and we observe relaxation. The same relaxation behavior is observed on the compressed fibers, superimposed by the visco-elastic relaxation of the Kelin-Voigt elements in the interval $t>25\tau$. The effects of reverse slip on the macroscopic behavior can be seen by following the dotted line for $t>15\tau$. After unloading, the non-slip strain vanishes, and the macrostrain is only formed by slip strain. 

\begin{figure}
        \centering
        \includegraphics[width=0.8\columnwidth]{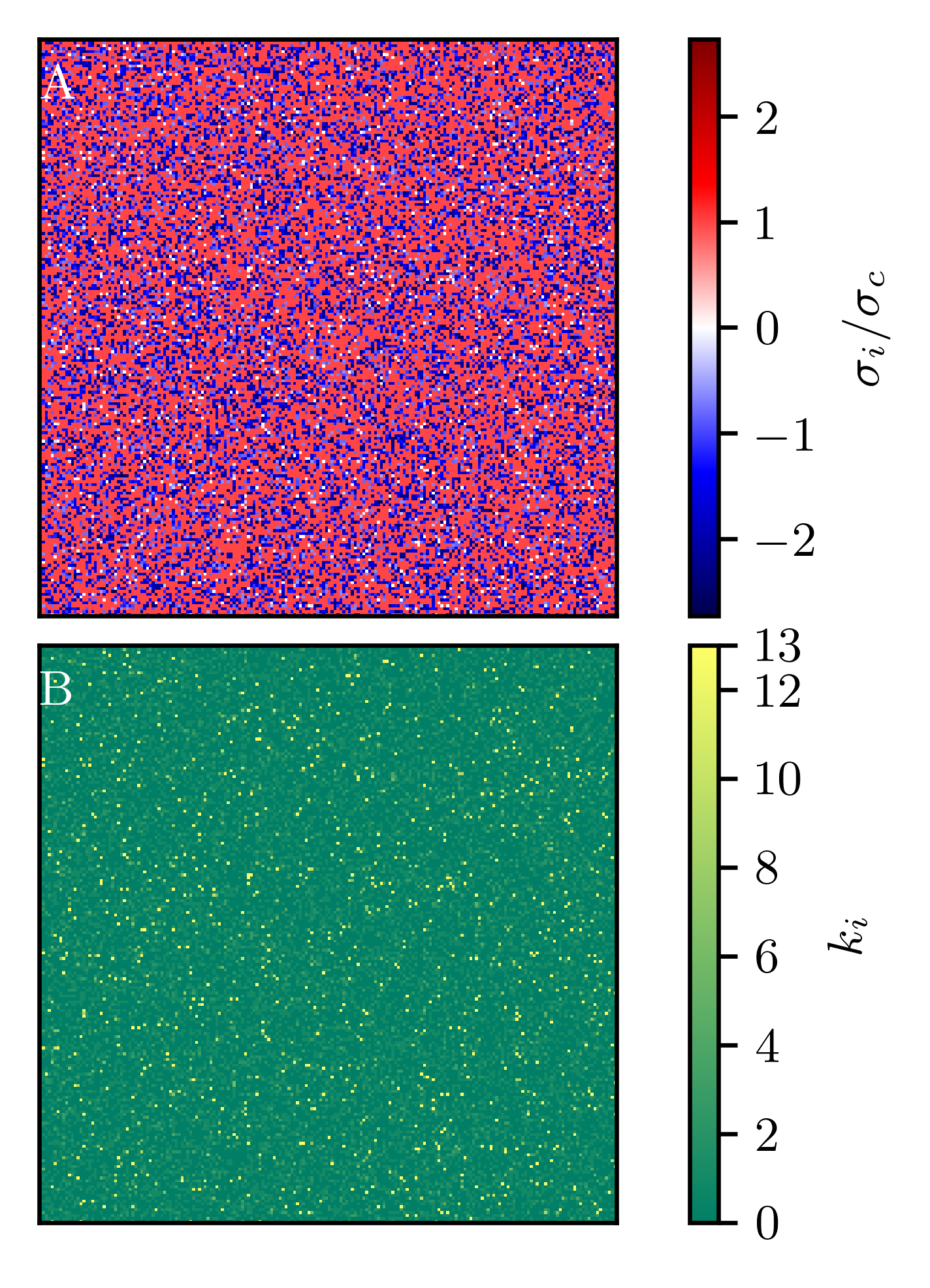}
        \caption{Residual normalized stresses (A) and number of slip events (B) for each fiber of the bundle at $t=40\tau$ in Fig.~\ref{fig:1cycle}.}
        \label{fig:maps}
    \end{figure}

Residual stresses develop in the system to balance the different slip strains of fibers. We show the residual stresses of each fiber at the end of the simulation with $D_w/D_d=1.3$ in Fig.~\ref{fig:maps}-A, and the matrix $k$, which counts the number of slips for each fiber, in Fig.~\ref{fig:maps}-B. By combining both figures, we see both of them with a predominant color: red (positive stresses) for Fig.~\ref{fig:maps}-A and green (low number of slips) for Fig.~\ref{fig:maps}-B. This shows that tensile stresses are stored on fibers with low slip numbers and, therefore, low slip strain. Fibers with higher slip numbers but still holding load are easily identified in Fig.~\ref{fig:maps}-A, being the ones with compressive stresses and marked in blue. Compressive stresses are observed on fewer fibers than the tensile ones; therefore, they are much higher (maximum $\sigma/\sigma_c\approx-2.7$) than tensile ones (maximum $\sigma/\sigma_c\approx1$). The fibers that reached the slip number $k^f$ are also easily identified in Fig.~\ref{fig:maps}-A.

%%%%%%%%
\subsection{Observations for multiple moisture-cycles}\label{Sec: Res: Many}
    \begin{figure*}[htb]
        \centering
        \includegraphics[width=\textwidth]{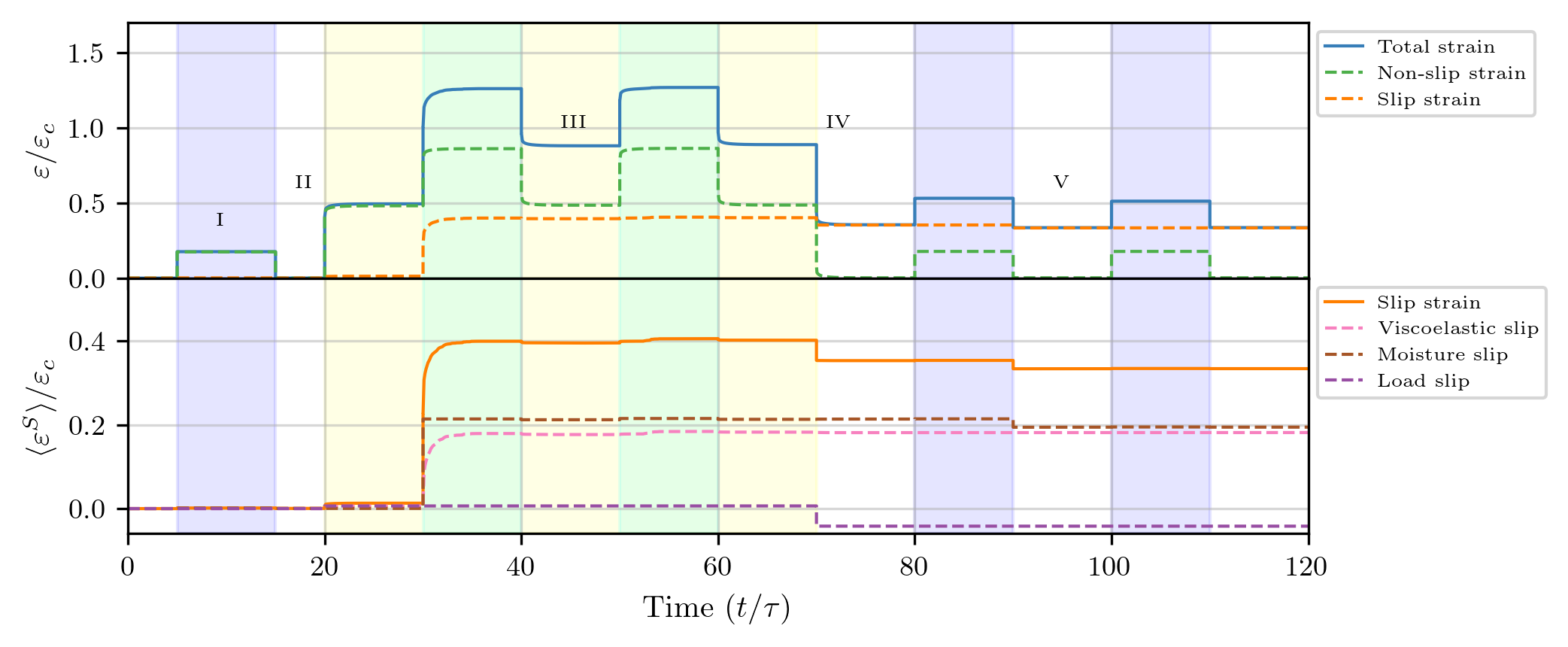}
        \caption{SS-FBM strain over five moisture cycles. Solid lines represent total values of $\varepsilon$ and $\varepsilon^S$ in both plots, respectively, while dashed lines represent their labeled components.}
        \label{fig:multicycles}
    \end{figure*}
The buildup of mechanosorptive strains and their relaxation in hygro-responsive materials requires loading histories with multiple MCs during load intervals and after load removal. To study these effects, we alternate the moisture within $10\tau$ lasting intervals for three different load cases: without load, within a constant load interval, as well as after unloading (see Fig.~\ref{fig:multicycles}). In the first MC (Fig.~\ref{fig:multicycles}, upper half), due to the increased strain, some slips occur (lower half), mainly from fibers with low thresholds from the tail of the Weibull distribution. Therefore, the slip strain is small. 

In the time interval $t=[20,50]\tau$, the situation of the single moisture cycle applies (Sec.~\ref{Sec: Res: Single}), but in the lower part of Fig.~\ref{fig:multicycles}, we now take a closer look into the evolution of the different components of the slip strain, split by their cause. At $t=50\tau$, a second MC is imposed on the loaded system, and again, like in the first MC under load, additional moisture and viscoelastic slips are triggered, resulting in an increased total strain. Each additional MC would further increase the total strain, since slip events, occurring during the cycle with increase and decrease, further reduce the slip thresholds even up to the failure of fibers. 

From $t=70\tau$ on, the system is unloaded, and elastic and viscoelastic relaxations occur just like in the single MC case. In each MC during this phase, frozen strains are relaxed in the system, mainly by moisture slips. Due to swelling at moisture increase, the compressed fibers are unloaded and therefore will not slip, while for moisture decrease, the compression of fibers results in reverse slip events, reducing the slip and consequently total strain. This effect is largest during the first relaxation cycle, but evidently, there exists a limit from where the total strain must be considered as irrecoverable. When we increase the number of MC in the loaded system, the SS-FBM can fail, which is the focus of the next study.
%%%%%%%%%%%%%%
\subsection{Failure analysis}\label{Sec: Res: fail}
\begin{figure}
        \centering
        \includegraphics[width=\columnwidth]{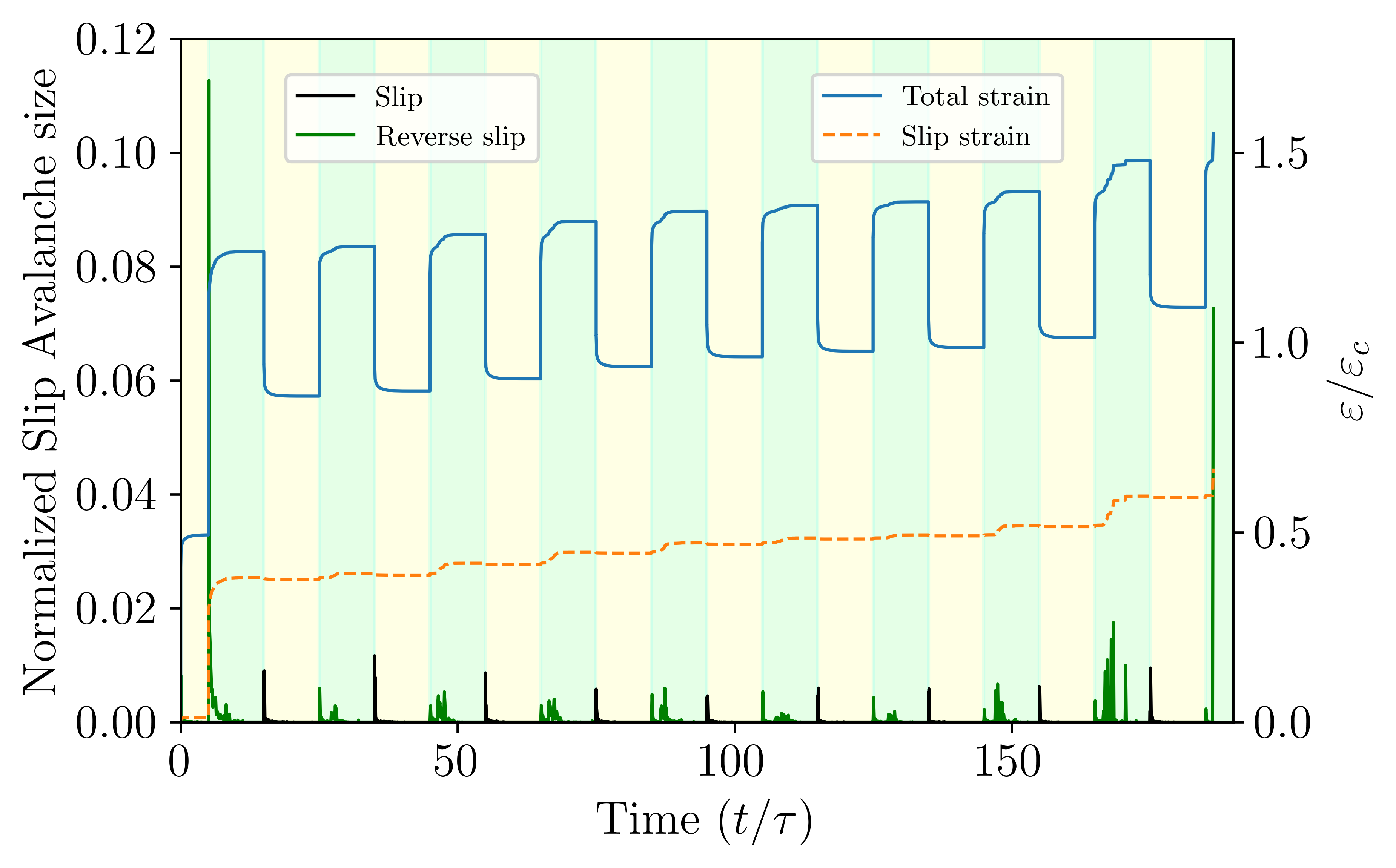}
        \caption{SS-FBM slip strain evolution over five moisture cycles and the respective normalized slip (green) and reverse slip (black) avalanches.}
        \label{fig:failure}
    \end{figure}

In this Section, we analyze how the cumulative slip avalanches, described in Sec.~\ref{Sec: Res: Many}, evolve up to the failure of the bundle. For this purpose, we plot the evolution of the normalized macrostrain $\varepsilon/\varepsilon_c$ and the normalized slip strain $\varepsilon^S/\varepsilon_c$ over time (see Fig.~\ref{fig:failure}). Additional insight is given by the slip avalanche sizes, normalized by the number of fibers ($N$), decomposed into slips and reverse slips. Two slip avalanches stand out in Fig.~\ref{fig:failure}: the one in the first MC and the one in the ultimate MC with the bundle failure. In between these two, both slip and reverse slip avalanches are comparable in size. However, the resultant strains of reverse slip avalanches are significantly smaller than the ones of slip avalanches. For one thing, slip avalanches can trigger fibers that have never slipped before and, therefore, have the highest slip threshold in the system up to that point. For another, reverse slip avalanches occur with fibers that have larger accumulated slip strains, which will have lower slip thresholds due to a higher value $k_i$. Finally, the reverse slip avalanches cannot relax the full slip strain obtained in the previous slip avalanche but are capable of weakening the bundle for the subsequent slip avalanche.

\begin{figure}
        \centering
        \includegraphics[width=\columnwidth]{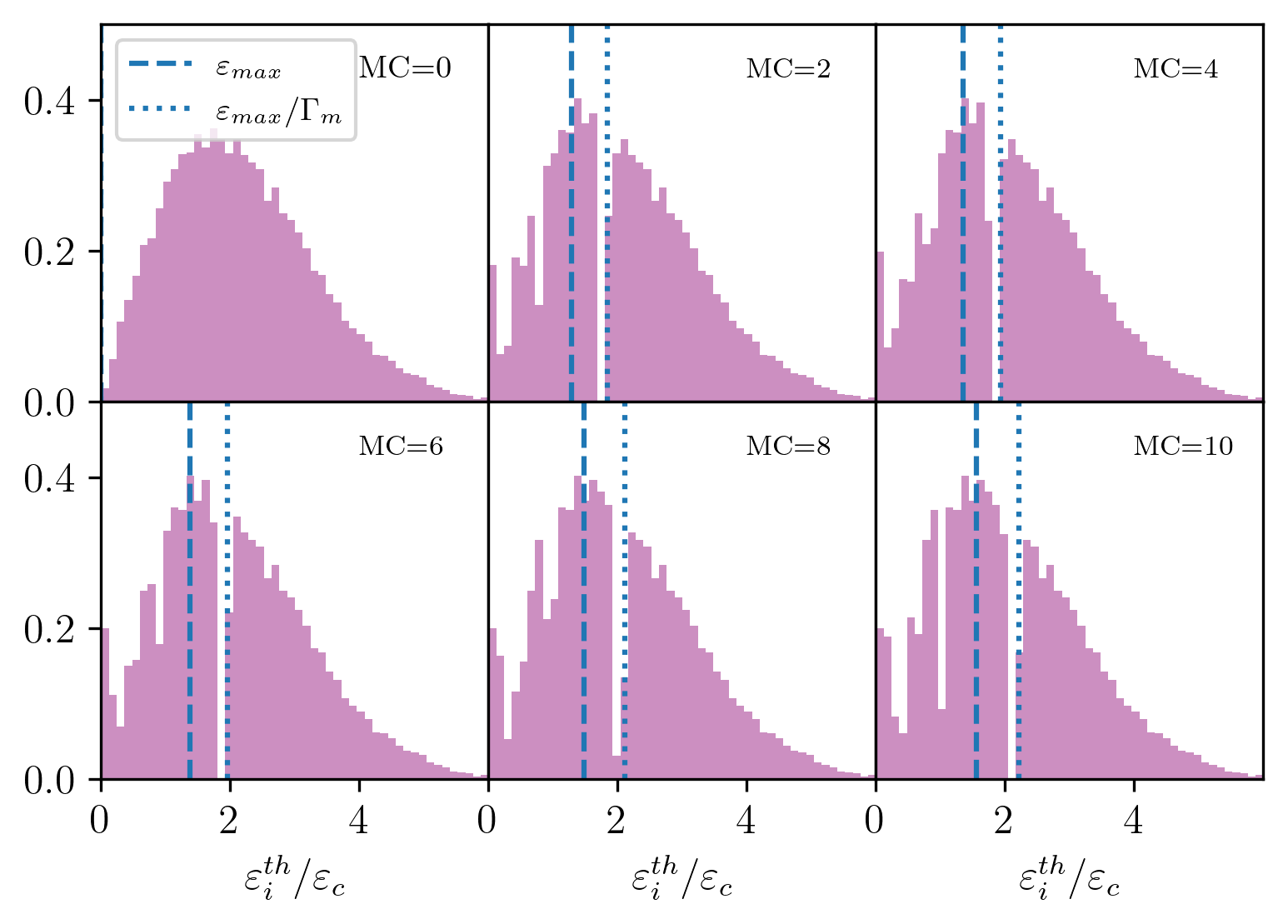}
        \caption{Probability density of slip thresholds for increasing moisture cycles MC with identical legend.}
        \label{fig:hist}
    \end{figure}

Moisture changes resemble a shift of the Weibull- threshold distribution, while the weakening by the function $f(\textbf{k})$ due to slips results in a degradation of the Weibull distribution. This effect is plotted for increasing moisture cycles in Fig. \ref{fig:hist}, starting with a typical Weibull distribution for $MC=0$. In the subsequent moisture cycles ($MC>0$), slip avalanches shift part of the distribution to lower values, resulting in a characteristic void. The void relates to the highest slip threshold triggered up to that point, and all slipped fibers will fill up the distribution on the lower side. Note, that due to the moisture scaling with factor $\Gamma_w$, we have two locations of maximal strains, one for the dry state (dashed line) and one for the wet one (dotted line). The ultimate degraded distribution (MC=10) is obtained before the last slip avalanche, representing the failure point. Failed fibers fill up the first bin of the distribution at $\varepsilon^{th}_i=0$. 

\begin{figure}
        \centering
        \includegraphics[width=1\columnwidth]{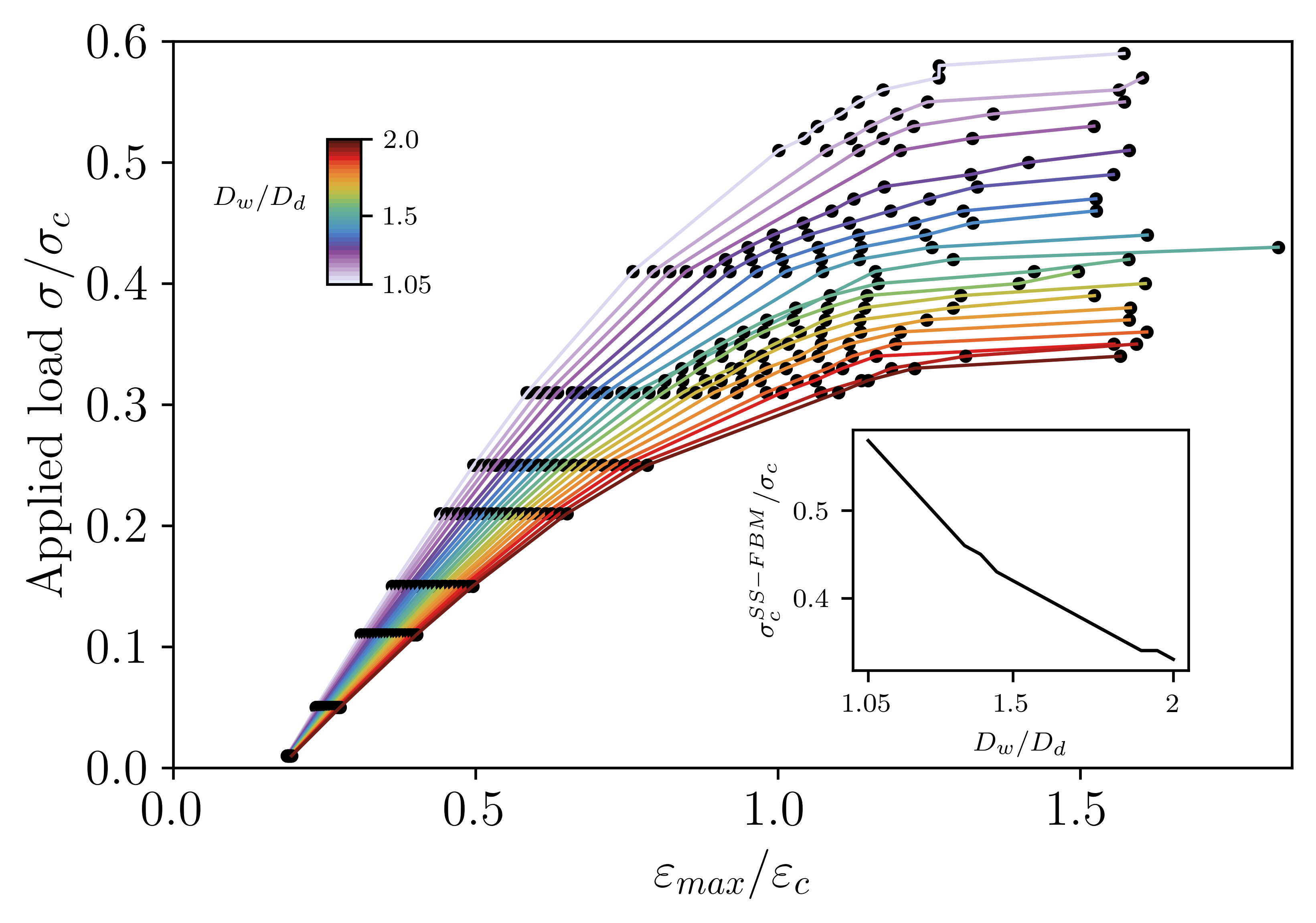}
        \caption{SS-FBM maximum strain-stress relation for different dry-to-wet compliance ratios and comparison to the strength with the dry FBM (inset).}
        \label{fig:strain_JwJd}
    \end{figure}

\begin{figure}
        \centering
        \includegraphics[width=\columnwidth]{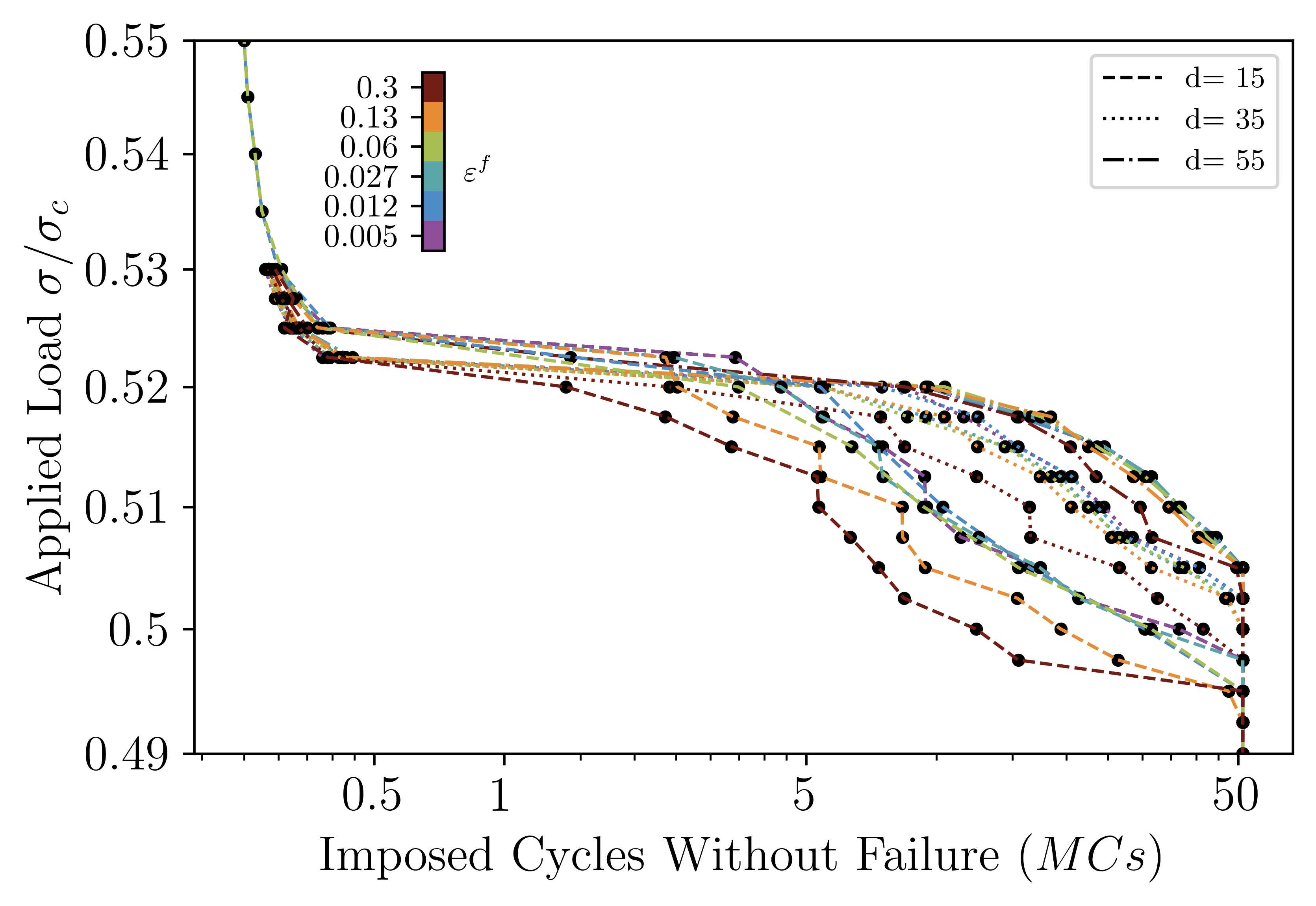}
        \caption{Applied load over the number of cycles that can be imposed before the failure of the system. If the system doesn't fail, the maximum number of imposed cycles is 50. }
        \label{fig:sigma_d}
    \end{figure}

Similar to typical experimental observations, the failure of our SS-FBM depends on intrinsic model properties and the loading history. We limit ourselves to exploring the effect of variations of the compliance ratio $D_w/D_d$ on the failure evolution (Fig.~\ref{fig:strain_JwJd}), as well as the effect of the weakening function with its decay parameter $d$ and failure limit $\varkappa^f$ on the number of endured moisture cycles up to failure (Fig.~\ref{fig:sigma_d}).

When changing the applied load $\sigma/\sigma_c$, we can observe the maximum normalized macrostrain $\varepsilon_{max}/\varepsilon_c$, from simulations, along with its strong dependence on the wet-to-dry compliance ratio $D_w/D_d$ (see Fig.~\ref{fig:strain_JwJd}). Each point in the plot resembles the maximum strain state of an individual simulation after 50 MCs. For low values of $\sigma/\sigma_c$ we observe a linear trend for $\varepsilon_{max}/\varepsilon_c$, independent of the $D_w/D_d$ ratio. The dependence of the strength ratio $\sigma_c^{SS-FBM}/\sigma_c$, meaning the strength ratio of the SS-FBM and the FBM, versus $D_w/D_d$ is given in the inset. As expected, the higher the moisture dependence of the compliance, the weaker the system, compared to the dry FBM.

Finally, we investigate the fatigue failure with respect to MCs, for varying loads and different values of the parameters of the weakening function. Compared to the variation of wet-to-dry compliances (Fig.~\ref{fig:strain_JwJd}), the strength is less sensitive when the weakening function is varied $f(k)$. Nevertheless, $d$ and $\varkappa^f$ affect the bearable number of MCs. Fig.~\ref{fig:sigma_d} shows the relation between applied load $\sigma/\sigma_c$ and the number of cycles imposed before failure for different combinations of $d$ and $\varkappa^f$. Combinations of values of $d=[15,35,55]$ and $\varkappa^f=[0.0055, 0.012, 0.027, 0.06, 0.13, 0.3]$ are shown. We closely examine the transition between failing after multiple MCs and failing within the first MC. It is interesting to observe that the increase of $d$ results in a more abrupt transition, which means a lower chance of failing due to the moisture cycles. Roughly, for higher values of $d$, the system either fails at the first moistening or doesn't fail at all. The second interesting observation is that the failure becomes less sensitive to variations on $\varkappa^f$ when $d$ increases. A high decay parameter $d$ means a lower weakening ratio, and as a consequence, it is more difficult to trigger strain increments during moisture cycles. As a result, the system's failure becomes less dependent on the failure criteria $\varkappa^f$, and the failure transition becomes more sensitive to load changes. In a limit situation, the load is high enough to fail the system in the first moisture cycle, or the system doesn't fail at all.

%%%%%%%%
\section{Conclusions}
We extended the stick-slip Fiber Bundle Model (SS-FBM) by~\citet{halasz2009fiber} with three additional features, namely the moisture-dependence of rheological elements, the possibility to slip under tension and compression, as well as history, moisture, and load direction dependence of slip thresholds. These three rules are inspired by wood anatomy~\cite{stevanic2020molecular} and observed hygro-mechanical behavior. We proved, that our modified moisture-responsive SS-FMB reproduces the main combined hygro-mechanical behaviors common to wood~\cite{armstrong1960effect,nordon1962some,gibson1965creep}. By demonstrating that this complicated response can be explained by just one mechanism, namely the slip, we open a new approach for exploring the structure-property relations in wood under combined hygric and mechanical load. 

From a fundamental modeling point of view, we altered the physical interpretation of slip strains by allowing their inversion. Instead of being plastic strains, therefore irreversible, we understand slip strains as frozen strains caused by rearrangements in the material structure with the potential for relaxation. Interestingly, no difference is made in whether the strain was developed by scleronomous stimuli, such as instantaneous load and moisture changes, or rheonomous ones, such as viscoelastic responses. More than that, the same source of non-linearity is responsible for the behavior observed in mechanosorption. In this way, complicated behaviors of hygro-responsive materials like plasticity, hygro-expansion, visco-elastic and mechanosorptive creep, which are typically described by different theories, are unified in just one simple model that embeds the concept of slip strains.

Most of these phenomena can already be seen in a single moisture cycle (MC) during loading with subsequent unloading (see Sec.~\ref{Sec: Res: Single}). For multiple MCs, the full descriptive potential of the SS-FBM becomes evident, as phenomena such as the build-up of mechanosorptive creep strains under combined load, as well as the relaxation of frozen strains during subsequent moisture cycles and creep plasticity naturally emerge. The numerical SS-FBM approach enables us to study the cyclic failure evolution and system behavior close to the failure point, based on the statistics of slip events and avalanches. The consequences of the history dependence of slip thresholds on the evolution of the threshold distribution, allow for a better interpretation of the damage progress of the system up to failure. Note that our assumptions on the effect of moisture, slip history and loading type are motivated by micro-structural considerations of the interaction of constituents in hygro-responsive (cellulose) materials. Our numerical study of the fatigue behavior gives insight into the role of these assumptions (see Sec.\ref{Sec: Res: fail}).

From a theoretical point of view, the chosen load sharing strategy offers an additional degree of variation. However, we stuck to democratic load sharing, since this most simple type of load sharing already produces the desired behavior, proving that locality in load redistributions is not a required feature here. The same can be said about a local degradation scheme for slipping fibers, even though micro-structural considerations would allow for such extensions, as preliminary simulations showed. In real systems, moisture gradients exist, that can result in stresses due to self-restrained swelling or shrinking, which additionally triggers slips on a time scale, determined by the moisture transport kinetics, calling for further investigation.

\section*{Acknowledgement}
The financial support from the Swiss National Science Foundation under SNF grant 200021\_192186 - Creep behavior of wood on multiple scales - is acknowledged. We acknowledge Prof. Dr. Ingo Burgert for the valuable discussions.

\bibliography{lit}% Produces the bibliography via BibTeX.

% \newpage

% \begin{align}
    
%     t=0=t_i \\
    
%     \text{While} t\textless t_{final} \rightarrow \text{"Time evolution loop"} \\
    
%     \text{Set} \sigma(t), \varphi(t) \\
    
%     \text{Compute} \varepsilon^H, \varepsilon \text{ (Eq.~\ref{epsilonmacro}), and } \boldsymbol{\varepsilon}_i^{th}(\boldsymbol{k},\varphi,\boldsymbol{\beta}) \text{ (Eq.~\ref{slipthesredcond})} \\
    
%     \text{While slip condition (Eq.~\ref{slip_cond})} \rightarrow \text{"Slip avalanche loop"}\\
 
%     \text{Define slip direction for each fiber} (\beta) \\
 
%    \text{Update } \boldsymbol{\varepsilon}_i^S, \varepsilon^S, \boldsymbol{k},\text{ and } \boldsymbol{\varepsilon}_i^{th}(\boldsymbol{k}) \text{ (Eq.~\ref{slipthesredcond})} \\
 
%     \text{Compute }\varepsilon \text{ (Eq.~\ref{epsilonmacro})}\\ 
 
%     \text{Time increment: } \Delta t = (t_{i+1}-t_i)/100, t^{*}=t_i\\
 
%     \text{Compute } \boldsymbol{\varepsilon}_i^{VE}(t^{*}\rightarrow t^{*}+dt) \text{ (Eq.~\ref{viscoelastic})}\\
 
%     \text{While slip condition (Eq.~\ref{slip_cond})} \rightarrow \text{"Slip avalanche loop"}\\
    
%     \text{Define slip direction for each fiber} (\boldsymbol{\beta}) \\
 
%     \text{Update } \boldsymbol{\varepsilon}_i^S, \varepsilon^S, \boldsymbol{k},\text{ and } \boldsymbol{\varepsilon}_i^{th}(\boldsymbol{k}) \text{(Eq.~\ref{slipthesredcond})} \\
 
%     \text{Compute }\varepsilon \text{ (Eq.~\ref{epsilonmacro})}\\  
 
%     t^{*}=t^{*}+dt
 
%     t_i=t_{i+i}   
 
%     \text{While } t^{*}\textless t_{i+1} \rightarrow \text{"Viscoelastic creep loop"}
    
% \end{align}

\end{document}